\documentclass[structabstract]{aa}  
\usepackage{graphicx}
\usepackage{txfonts}
\begin{document}
   \title{The P Cygni supergiant [OMN2000] LS1 - implications for  the star formation history of W51}
   \author{J. S. Clark\inst{1}
\and B. Davies\inst{2,3}
\and F. Najarro\inst{4}
\and J. MacKenty\inst{5}
\and P. A. Crowther\inst{6}
\and M. Messineo\inst{3}
\and M. A. Thompson\inst{7}
}
\institute{
$^1$Department of Physics and Astronomy, The Open 
University, Walton Hall, Milton Keynes, MK7 6AA, UK\\    
$^2$School of Physics and Astronomy, University of Leeds, Leeds, LS2 9JT, UK \\
$^3$Chester F. Carlson Centre for Imaging Science, Rochester Institute of Technology, 54 Lomb Memorial 
Drive, Rochester NY 14623, USA\\ 
$^4$Departamento de Astrof\'{\i}sica, Centro de Astrobiolog\'{\i}a, (CSIC-INTA),
Ctra. Torrej\'on a Ajalvir, km 4,  28850 Torrej\'on de Ardoz, Madrid, Spain\\
$^5$ Space Telescope Science Institute, 3700 San Martin Drive, Baltimore, MD 21218, USA\\
$^6$Department of Physics and Astronomy, University of Sheffield, Hounsfield Road, Sheffield, S3 7RH, UK\\
$^7$Centre for Astrophysics Research, University of Hertfordshire, College Lane, Hatfield, AL10
9AB, UK} 

   \abstract{}{We investigate the nature of the massive star \object{[OMN2000] LS1}
    and  use these results to constrain the history of star formation within the host complex W51.}
{We utilised a combination of near-IR spectroscopy and non-LTE model atmosphere analysis to derive
the physical properties of \object{[OMN2000] LS1}, and a combination of theoretical evolutionary calculations and 
Monte Carlo simulations to apply limits on the star formation history of W51.}
{We find the spectrum of \object{[OMN2000] LS1} to be consistent with that of a P Cygni supergiant. 
With a temperature in the range of 13.2-13.7kK and log($L_{\ast}/L_{\odot}$)$\leq$5.75,
it is significantly cooler, less luminous, and less massive than proposed by previous authors. 
The presence of such a star within 
W51  shows that star formation has been underway for at least 3Myr, while  the formation of massive O stars is still 
on going. The lack of a population of evolved red supergiants within the complex shows that the rate of formation of young massive clusters at 
ages $\geq$9Myr was lower than currently observed. We find no evidence of internally triggered,
 sequential star formation within W51, and favour the suggestion  that star formation has proceeded at
 multiple indepedent sites within the GMC. Along with other examples, such as the G305 and Carina star-forming regions, we suggest that W51 is a Galactic analogue of the ubiquitous star cluster complexes seen in external galaxies such as M51 
and NGC2403.}
{} 


   \maketitle

\section{Introduction}

While the physical mechanism for building 
 massive ($>$20M$_{\odot}$) OB stars remains hotly debated, it 
appears likely that the majority of these stars form in star clusters 
(de Grijs \cite{deG}, Parker \& Goodwin \cite{parker}) rather than isolation. 
Moreover, observations of nearby galaxies reveal that such clusters 
 form in larger complexes  (e.g. M51; Bastian et al. 
\cite{bastian}). 
Unfortunately, the process(es) that  converts
giant molecular clouds (GMCs) into such complexes and the timescale for their formation currently 
 remain opaque. This in part is a consequence of the  restricted spatial resolution of
 extragalactic  studies, which  compromises the determination of both stellar and cluster ages and 
 mass functions.

Consequently, it is instructive to  search for Galactic analogues of star-forming
complexes that may be observed with enough resolution that individual 
(proto-)stars may be studied. First detected by Westerhout 
(\cite{westerhout}), W51 consists of two giant H\,{\sc ii} regions, 
 \object{W51A} \& \object{W51B}, both of which  may in turn be resolved into 
smaller components (e.g. Mehringer \cite{mehringer}, Nanda Kumar et al \cite{nanda} and references 
therein). 
With an angular extent of  1$^{\rm o} \times 1^{\rm o}$, and a mass of $\sim 10^6$M$_{\odot}$, W51 is amongst the most massive Galactic GMCs (Carpenter \& Sanders \cite{carpenter}),
while the giant H\,{\sc ii} regions  imply a large 
population of O stars to yield the requisite UV ionising flux; it  therefore
represents an excellent candidate  for a massive star formation (SF) complex. 

Near-IR  imaging of W51 by Okumura et al. (\cite{ok}; OMN2000) and  Nanda Kumar et 
al. (\cite{nanda}) indicated a significant population of young 
O  stars and massive young stellar objects (YSOs). Subsequent spectroscopic observations 
by Figueredo et al. (\cite{figueredo}) and Barbosa et al. (\cite{barbosa}) confirmed these 
findings, identifying several early-mid O stars within W51A and resolved the 
subregion W51 IRS2 (see Fig. 1) into a proto cluster containing  an $\sim$O3 star and a 
massive YSO. With the recent results of Zapata et al. (\cite{zapata}) revealing
the possible formation of a further massive ($\sim$40M$_{\odot}$) proto star in the nearby 
region W51 North, it is clear that vigorous SF is currently underway within  W51A.

In order to investigate the stellar content and  SF history of W51 in detail 
we embarked on a comprehensive spectroscopic and imaging survey of the complex from near-IR to radio wavelengths. Here we present the first 
results of this investigation,  focusing on the source  \object{[OMN2000] LS1}, 
located at one extremity of the complex (Fig. 1; RA=19 23 47.64 $\delta$=+14 36 38.4). 
 OMN2000 described  it as a `P Cygni type supergiant' and classified it 
as O4 I, with a progenitor mass of $\sim$120M$_{\odot}$, making it of considerable interest as 
one of the most massive, evolved stars in the Galaxy. In this work we present new high 
resolution spectroscopy of this object which allow us to better constrain its stellar 
properties and discuss the implications of these results for the distance to, and SF history 
of, W51.

\begin{figure*}
\resizebox{\hsize}{!}{\includegraphics[angle=-90]{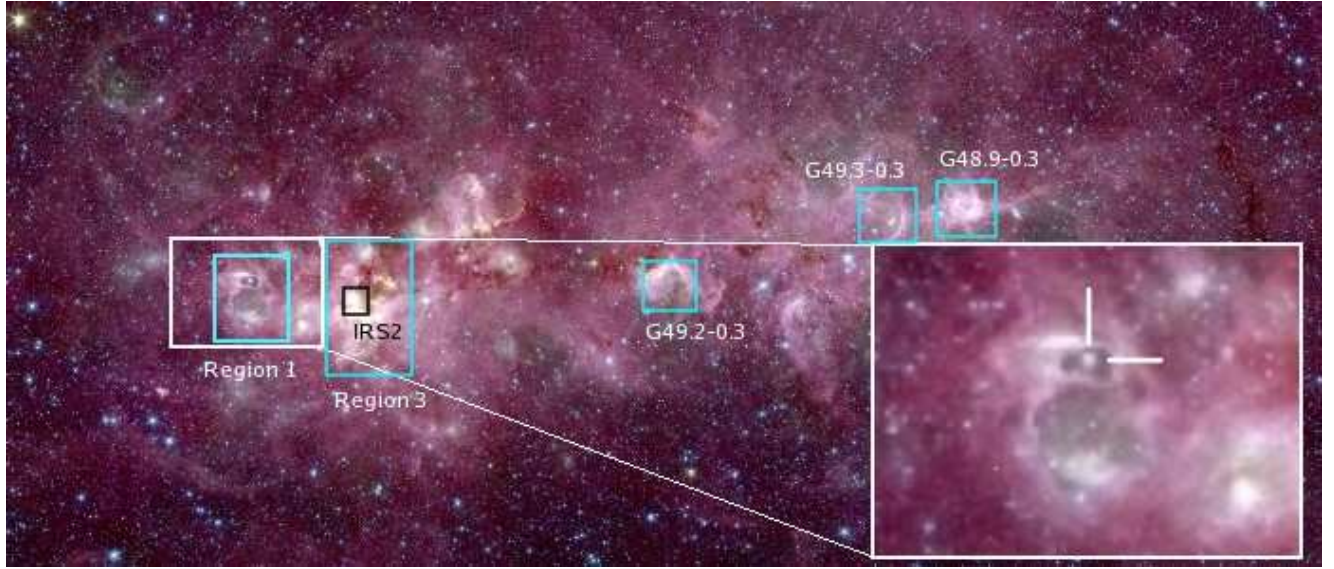}}
\caption{Spitzer 3 colour image of W51 (blue - 3.6$\mu$m, green - 4.5$\mu$m \& red - 8.0$\mu$m) 
with the position of \object{[OMN2000] LS1} indicated. The overal length of the W51 complex is 
$\sim$100(d/6kpc)pc, with the apparent wind blown bubble surrounding \object{[OMN2000] LS1} 
being $\sim(1.8\times1.0)$(d/6kpc)pc. For clarity we also indicate SF regions 1 \& 3 as defined by OMN2000 
(the latter containing the O star candidates identified by Figueredo et al. \cite{figueredo}), the heavily
 embedded massive YSO forming
complex IRS2 and 3 further regions throughout the W51 complex  which Nanda Kumar et al. 
(\cite{nanda}) find to host massive young stellar clusters (the fourth region studied in detail by 
Nanda Kumar et al. encompasses IRS2 and is located within Region 3 of OMN2000). 
Note that for reasons of clarity not all the locations of star formation activity 
identified by these authors are indicated in this figure.}
\end{figure*}

\section{Data Reduction \& Presentation}

Data was taken during the night of 27 September 2007, using the
Infra-Red Multi-Object Spectrograph (IRMOS, MacKenty et al. \cite{mac})
mounted on the Mayall 4-m at the Kitt Peak National Observatory,
 with a 1.9 to 2.2 micron (K1) band pass filter with a grating providing l/dl 
of 3000. This gave us a resolution of $\sim$100kms$^{-1}$ in the the spectral range
2.0--2.2microns.
To correct for the variability in the remaining background signal, we
followed each science integration with a dark-frame of equal
integration time. To compensate for variable sky
background, we limited our science exposures to 2min. The star was
dithered along the slit by 2arcsec every five science exposures to
compensate for artifacts on the detector. In total, we integrated on
the object for 1 hour, with 30 individual science exposures
All data reduction was done using custom-written routines in IDL.
Initially, each science frame had the dark frame taken closest in time
subtracted from it. The science frames were then coadded, and divided
through by the normalized flat-field.

Correction for geometric distortion in both  the spatial and dispersion
directions are required prior to spectral extraction. To accomplish this, 
the data were resampled onto a linear grid. The spatial warping was characterized by
fitting a 3rd degree polynomial through the spectral traces of the
star in each of the dither positions. The warping in the dispersion
direction was determined by linear fits to the OH emission lines in
the sky either side of each spectral trace. Using the fits to the
stellar spectral trace and the sky lines as tie-points, the warping
was fitted using a two-dimensional, third-degree polynomial. The
inverse of this 2-D fit was then applied to the data to resample it
onto a linear grid. As the wavelengths of the sky OH lines are known,
the data is wavelength-calibrated in this de-warping process.
The spectrum was optimally-extracted from the de-warped data using the
algorithm of Horne (\cite{horne}).

To remove the atmospheric absorption, the object spectrum was divided
through by that of a telluric standard (SAO~107138, spectral type A8 V). Prior to 
division, the two spectra were first cross-correlated to correct for any sub-pixel
shifts which would produce artifacts in the final spectrum. Finally,
the data were normalized by the mean continuum value, and the resultant spectrum 
plotted in Fig. 2.

\section{The Nature of \object{[OMN2000] LS1}} 

\begin{figure}
\resizebox{\hsize}{!}{\includegraphics[angle=90]{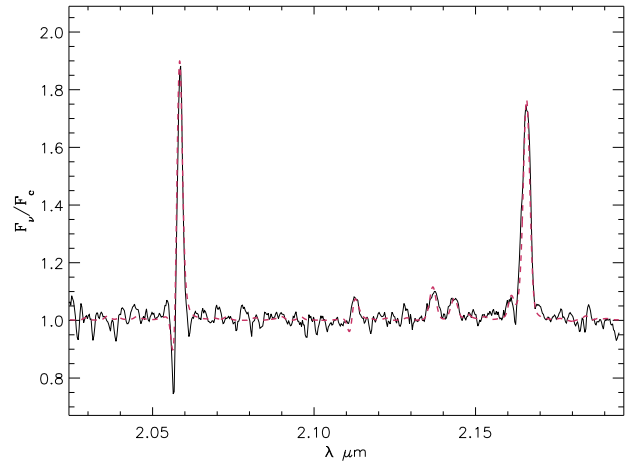}}
\caption{High resolution K band spectrum of \object{[OMN2000] LS1} (solid line) with 
the best model fit superimposed (red dashed lines). The major transitions present
are He\,{\sc i} 2.058$\mu$m \& 2.1128$\mu$m, Br$\gamma$, and Mg\,{\sc ii}
 2.138/44$\mu$m.}
  \end{figure}

Comparison of the low resolution spectrum of OMN2000 to ours reveals no obvious changes 
in a decade. Both spectra are dominated by He\,{\sc i} 2.058$\mu$m \& 2.112$\mu$m, 
Br$\gamma$ and Mg\,{\sc ii} 2.138 \& 2.114$\mu$m emission.  However, the improved resolution
 and S/N($\sim$70) of our observations now clearly reveals the  presence of a P Cygni profile
 for  He\,{\sc i} 2.058$\mu$m that was hinted at in the spectrum of OMN2000; weak emission 
in He\,{\sc i} 2.161 \& 2.162$\mu$m likely  veil a similar feature in Br$\gamma$. 
The formal classification of O4 I by OMN2000 is incorrect, due to a lack of  high 
excitation species such as N\,{\sc iii} and C\,{\sc iv}
(Hanson et al. \cite{hanson96}). However,  the qualitative description as a `P Cygni 
supergiant' is more acccurate, with \object{[OMN2000] LS1} bearing a close resemblance to 
known examples (which include confirmed  and   candidate Luminous Blue Variables (LBVs):
 Clark et al. \cite{clark03}, Voors et al. \cite{voors}). 
While  a P Cygni profile in He\,{\sc i} 2.058$\mu$m is not seen in all such stars, it is present in 
 \object{HD316285} (Hillier et al. \cite{hillier}) and the Galactic Centre Ofpe/WN9
stars (Najarro et al. \cite{naj97}; also known as WN9-11h stars) which have spectra similar to known LBVs such as AG Car in the hot state.
We therefore conclude that the spectrum of   \object{[OMN2000] LS1} is consistent with a a qualitative 
classification  as a P Cygni-type B  supergiant (and hence potentially a LBV).

\subsection{The distance to \object{[OMN2000] LS1} and W51}

To derive the stellar parameters for \object{[OMN2000] LS1} one must adopt a distance to it and, 
by extension, the W51 complex. Figueruedo et al. (\cite{figueredo}) provides a summary of  the results of previous studies. 
These show that the  kinematic estimate  of the distance - utilising  radio recombination  lines - of 5.5kpc (Russeil \cite{russeil}),  is  broadly comparable to the results derived  from   maser proper motion measurements; 
6.1$\pm$1.3kpc (Imai et al. \cite{imai}) to 8.5$\pm$2.5kpc (Schneps et al. \cite{schneps}).
However, utilising 4 O stars which they classify as (Zero Age) Main Sequence objects, Figueruedo et al. (\cite{figueredo}) 
report a significantly smaller distance to W51 of 2.0$\pm$0.3kpc. Subsequnetly, 
Barbosa et al. (\cite{barbosa}) suggest an upper limit of 5.8kpc 
by equating the radio luminosity of IRS2 with the ionising flux from the O star W51d under the assumption that it too is a
 Main Sequence star. Finally and most recently, Xu et al. (\cite{xu}) report a 
distance of 5.1$^{+2.9}_{-1.4}$kpc based on trigonometric parallax measurements.

Both  Figueredo et al. (\cite{figueredo}) and Xu et al. (\cite{xu}) recognise the difficulty in reconciling the spectroscopic 
distance with the other estimates but are unable to provide an explanation for this discrepancy. Possible reasons for 
underestimating the spctroscopic distance would be the adoption of an incorrect reddening law to W51, unrecognised 
binarity or multiplicity in the stars or an incorrect spectral classification. Relating to the final point we note that
 the luminosity of early O stars may be determined from the Br$\gamma$ line (Hanson et al. \cite{hanson05}),
 with supergiants demonstrating infilling or emission. While the emission observed in three of the four stars studied by
 Figueredo et al. (\cite{figueredo}) may result from incomplete nebular subtraction, if the stars were Main Sequence objects, 
the stark absorption wings in the Br$\gamma$ profile would be visible given the S/N and resolution of the spectra. Thus we suggest that the  luminosity class of these stars - adopted due to their proximity to a star forming region - and hence their distance may be underestimated.

 For  a distance to W51  of 6kpc, the O stars discussed by Figueredo et al. (\cite{figueredo}) are at projected 
distances of 2-4pc from the compact star forming region IRS2. While it might appear  unlikely for 
evolved stars to be located so close to regions of ongoing SF, early O supergiants are 
found  within the massive star forming regions G305 (Leistra et al. \cite{leistra})
and  W43 (Blum et al. \cite{blum}). Thus non-coevality of the stellar population(s) may be a common feature 
of large star forming complexes such as W51 and G305 (Davies et al., in prep.). 
Nevertheless, given this uncertainty, we undertook our analysis of 
 \object{[OMN2000] LS1} for both near, spectroscopic (2kpc) and far, kinematic and parallactic distances (for which we adopted a representative value of  6kpc)
as well as for a third, intermediate (3.4kpc) distance, the choice of which is justified in Sect. 3.3.

\subsection{Physical parameters of \object{[OMN2000] LS1}}

We have used CMFGEN, the iterative, non-LTE line blanketing method
presented by Hillier \& Miller (\cite{hil98}) to model \object{[OMN2000] LS1} and estimate the physical
properties of the star. The method solves the radiative transfer equation in the
co-moving frame and in spherical geometry for the expanding atmospheres
of early-type stars. The model is prescribed by the stellar radius,
$R_{\ast}$, the stellar luminosity,
$L_{\ast}$, the mass-loss rate $\dot M$, the velocity field, $v(r)$ (defined by
$v_{\infty}$ and $\beta$), the volume filling factor characterizing the clumping
of the stellar wind, {\it f(r)}, and elemental
abundances. Hillier \& Miller (\cite{hil98},\cite{hil99}) present a detailed 
discussion of the code.

For the present analysis, we have assumed  solar metalicities for an atmosphere composed of
H, He, C, N, O, Mg, Si, S, Fe and Ni. Observational constraints are
provided by the K-band spectrum and the 2MASS J, H and K
photometry.  The validity of our technique has
been demonstrated in Najarro et al. (\cite{naj99}) and Najarro (\cite{naj01}) 
by calibrating our method against stars with similar spectral type such as P~Cygni and
HDE~316285 (for which optical and UV spectra
are also available). We refer to Najarro et al. (\cite{naj08}) for a detailed review on the
analysis technique and present the  results from fits for the three distances adopted in Table 1.
As described by Najarro et al. (\cite{naj08}), 
the computationally intensive analysis precludes sufficient numbers of models  being calculated to permit 
statistically robust error estimates to be made for all parameters. Consequently, the 
uncertainties presented in Table 1 represent the range of values for which an acceptable fit to the 
available data is acheived, an approach also adopted by Martins et al. (\cite{martins07}).

The primary diagnostics used were the presence and  relative strengths of the three  He\,{\sc i} emission lines at 2.058$\mu$m, 2.112$\mu$m, 2.185$\mu$m, the He\,{\sc i} (7-4) complex around 2.16$\mu$m, Br$\gamma$ and the Mg\,{\sc ii}  lines (2.138$\mu$m and 2.144$\mu$m). A lower temperature 
solution may be excluded due to the lack of any Fe\,{\sc ii} semi-forbidden lines (seen in
emission for cooler P Cygni supergiants; Geballe et al. \cite{geb00}), while the He\,{\sc i} 2.058$\mu$m line 
would also be expected to be significantly weaker. Conversely,  higher temperatures would lead to
 significantly stronger emission in this line - both in absolute terms and also in  relation to Br$\gamma$ - 
than is observed. 

P Cygni supergiants with high mass loss rates potentially demonstrate a  degeneracy between the H/He 
ratio and  mass loss rate. This occurs when the combination of stellar temperature and high wind density 
causes He\,{\sc ii} to recombine  to He\,{\sc i} very close to the photosphere. 
In such a situation any He/H ratio may fit the observations  with an appropiate scaling of the mass loss rate and a small variation  of the stellar temperature;  this effect is observed in  e.g.  \object{HD316285}
 (see Hillier et al. \cite{hillier} for a  discussion). However, unlike this star, the  lower mass loss rate 
 for  \object{[OMN2000] LS1} leads to a reduced wind density which, when combined with a higher effective temperature,
 minimises the effect of this degeneracy.  H/He ratios of 0.5 to 1.5 are permitted by our modeling; 
if H/He were higher then the He\,{\sc i} 2.112 and 2.185$\mu$m lines and He\,{\sc i} (7-4) complex would 
be weaker than observed. Following the scaling determined by Hillier et al. (\cite{hillier}) for HD316285, 
these abundances correspond to  a reduction in the mass loss rate of $\sim$60\% for H/He=1.5 when 
compared to that determined for H/He=0.5 (Table 1) but only  a small change ($<$300K) in temperature. 
For this  parameter regime, $\tau \sim$2/3 is already reached at the base of the wind, and so no information
is obtained from the hydrostatic layers of the star, the strong wind consequently 
fully determining T$_{eff}$ (which is defined at $\tau \sim$2/3).
 
The value of the reddening presented in Table 1 was determined  via comparison of the predicted
stellar near IR colour to 2MASS photometry of the source in order to determine the near IR excess due to reddening, 
 E(J-K). A value of A$_K$ was then calculated via the relation A$_K$=0.67E(J-K) and finally converted to 
E(B-V) via the relations 0.112A$_V$=A$_K$ and A$_V$=3.1E(B-V) (Rieke \& Lebofsky \cite{rieke}). The random error 
for the observed (J-K) from the 2MASS data is small ($\pm$0.03mag), as is the systematic 
uncertainty in the intrinsic (J-K)$_{\rm o}$ colour of the star (due to the  range of mass loss 
rates permitted by the modeling); we conservatively estimate an uncertainty in  A$_K$ of $\pm$0.1mag. 
The well developed P Cygni profile in He\,{\sc i} 
2.058$\mu$m  permits a determination of the terminal velocity of the wind ($v_{\infty}$) of $\pm$50kms$^{-1}$, 
Finally, we assume an uncertainty in log(L$_{bol}$) of $\pm$0.1dex, which we regard as highly conservative,
 given the small systematic uncertaintites in both  reddening and the temperature dependant bolometric correction.

\begin{table*}
\begin{center}
\caption{Derived stellar parameters for \object{[OMN2000] LS1} as a function distance (Sect. 3.1 \& 3.3).}
\begin{tabular}{cccccccccccccc}
\hline
Distance & $R_{\ast}$     & $T_{eff}$ & log($L_{\ast}/L_{\odot}$) & H/He & $v_{\infty}$ & $\beta$ &
{\it f(r)} & $\dot M$ & E(B-V) & A$_K$ & M$_K$ & M$_{initial}$ \\
 (kpc) & ($R_{\odot}$)& (kK)      &                  &      & (kms$^{-1}$) &      &
           & log($M_{\odot}$/yr) &    & & & (M$_{\odot}$)   \\
\hline
6  & 145.0 & 13.2 & 5.75 & 0.5 / 1.5  & 400 & 3.0 &0.08 & -4.2 / -4.6 &3.5 & 1.2  &-8.90 & $\sim$40\\
3.4&  82.5 & 13.4 & 5.30 & 0.5 / 1.5  & 400 & 3.0 &0.08 & -4.6 / -5.0 & 3.5 & 1.2 &-7.65 & $\sim$25\\
2  &  48.0 & 13.7 & 4.86 & 0.5 / 1.5  & 400 & 3.0 &0.08 & -4.9 / -5.3 &3.5 &  1.2 &-6.50& $\sim$17\\
\hline
\end{tabular}
\end{center}
 {As described in Sect. 3.2, H/He ratios suffer from a modest degeneracy, resulting in a reduction in 
the  mass loss rate by $\sim$60\% for the   H/He=1.5 model compared to H/He=0.5.  We further estimate systematic 
uncertainties of  $\pm$50kms$^{-1}$ in $v_{\infty}$, $\pm$300K in $T_{eff}$ and $\pm$0.1dex 
in log($L_{\ast}/L_{\odot}$).}
\end{table*}

\subsection{The evolutionary state  of  \object{[OMN2000] LS1}}

Based on the near-IR magnitudes of \object{[OMN2000 LS1]} and the  properties of the wind blown bubble surrounding it 
(H\,{\sc ii} region j), OMN2000 propose $T_{eff} \sim$40kK and log($L_{\ast}/L_{\odot}$)$\sim$6.3, yielding an initial mass 
of $\sim$120M$_{\odot}$. Irrespective of whether  a distance of 2 or 6kpc is correct, we find these values to be significant overestimates. We may also use our results 
to discriminate between the two distance estimates on the grounds of self consistency from both observational and 
theoretical perspectives. 

Firstly, the K band spectrum  suggests a  classification as a P Cygni supergiant. If it were 
located at 2kpc, \object{[OMN2000 LS1]} would have a luminosity  $\geq$0.4 dex below
   the lower end of the range observed for such stars (log($L_{\ast}/L_{\odot}$)$\sim$5.3 as seen for \object{HD~168625};
 Clark et al. \cite{clark05}). In contrast  the resultant  luminosity at 6kpc is  entirely consistent with such a spectral
classificaton. Moreover the other physical parameters such as mass loss rate, terminal wind velocity and H/He ratio 
are also fully consistent with those derived from non-LTE modeling of other P Cygni-type supergiants
  such as \object{P Cygni}, \object{HD 316285} (Hillier et al. \cite{hillier}) and \object{AG Car}
 (Groh et al. \cite{groh})  and the closely related WN9-11 stars (e.g. Najarro et al. \cite{naj97},
 Martins et al. \cite{martins07}).

With the temperatures and luminosities given in Table 1, comparison to the evolutionary tracks of  
 Meynet \&  Maeder (\cite{meynet}) suggest initial masses in the region of $\sim$40M$_{\odot}$ and $\sim$17M$_{\odot}$
for 6kpc and 2kpc respectively (Table 1). While it is thought that stars of $\geq$25M$_{\odot}$ will encounter a P Cygni supergiant phase during their post-red supergiant (RSG) bluewards evolution across the HR diagram, it is not expected that lower mass stars will evolve in such a manner, instead exploding as SNe while  RSGs, again favouring a larger distance. 

We may test this theoretical prediction via observations of the stellar populations of young, massive coeval 
clusters. Several  P Cygni supergiants  have  been identified in Westerlund 1 ($\sim$4-5Myr; 
Clark et al. \cite{clark05a}) and the Galactic  Centre ($\sim$6Myr; Najarro et al. \cite{naj97},
Martins et al. 
\cite{martins07}, Paumard et al. \cite{paumard}), Quintuplet ($\sim$4$\pm$1Myr; Figer et al. \cite{figer99}) 
 and 1806-20 ($\sim$3-5Myr; Figer et al. \cite{fig1806}, Bibby et al. \cite{bibby}) clusters. In all cases
 the stellar contents and ages of these clusters are  consistent with progenitor masses for these stars 
in the $\sim$30-60M$_{\odot}$ range. 
In contrast, despite a sample size of 30 clusters  which are massive enough to host blue and/or red supergiants  
and have  ages  in excess of 10Myr (Eggenberger et al. \cite{egg}, Davies et al. \cite{davies07}, \cite{davies08})
- appropiate for the post-Main Sequence (MS) evolution of stars of 
$\leq$20M$_{\odot}$ -  to the best of our knowledge to date no  P Cygni supergiant has been 
identified within any of them\footnote{Note that the intrinsic luminosity and strong emission line spectrum of 
a P Cygni supergiant would identify them in either a spectroscopic or a broad+narrow (e.g. H$\alpha$) band imaging survey.}.
This is even  true for such massive, well stocked clusters as RSGC1 \& 2 
(Davies et al. \cite{davies08}) and we therefore conclude that stars of such relatively low masses do not pass 
through such a phase at solar metalicities.

Thus, while we cannot  {\em a priori} exclude the possibility that  the  P Cygni supergiant phase occurs at 
sufficiently  low ($\leq$20M$_{\odot}$) masses to be reconciled  with a distance to \object{[OMN2000 LS1]} of 2kpc,
 we suggest that a distance of $\sim$6kpc is more consistent with current observational {\em and} theoretical 
constraints on the nature of such stars {\em and} the majority of current distance estimates for W51, including the 
recent parallactic estimate  of Xu et al. (\cite{xu}). 
Finally,  requiring the luminosity of \object{[OMN2000 LS1]} to match that of the faintest P Cygni supergiant
identified to date (log($L_{\ast}/L_{\odot}$)$\sim$5.3; see above) results in a  distance of 3.4kpc
(Table 1). Such a luminosity would correspond to an initial stellar mass of $\sim$25M$_{\odot}$ - a value on the 
cusp of that required by theoretical predictions for a star to evolve bluewards after exiting the RSG phase and 
hence potentially become a P Cygni supergiant\footnote{Indeed two clusters in the  Large Magellanic Cloud are found to contain the closely related 
 Ofpe/WN9 stars; LH39 ($\sim$8-10Myr) and LH89 ($\sim$5-7Myr; both ages from 
Massey et al. \cite{massey}). Thus it appears that under certain conditions 
stars with masses as low as $\sim$25M$_{\odot}$ may become P Cygni supergiants.}.

We conclude  that \object{[OMN2000] LS1} is an evolved massive star in  a P Cygni supergiant phase, with a high mass loss rate and evidence for chemical enrichment. While such a classification does not premit an exact distance to be determined, we favour a distance compatable with recent kinematic and parallactic 
estimates to W51, which would imply a progenitor mass of the order of $\sim$40M$_{\odot}$.

\section{The star formation history of  W51}

Combined with the identification of massive protostars, the presence of \object{[OMN2000] LS1} within W51A 
indicates that massive star formation has been underway for a
 significant  period of time. OMN2000  report  ages of only 1.9~Myr and 2.3~Myr for \object{[OM2000] LS1} and 
the region of W51 which hosts it (their 'Region 1'; Fig. 1). However, comparison of our modelling
results to the theoretical predictions of (Meynet \&  Maeder \cite{meynet}) suggest this is an 
underestimate, with the episode of SF that yielded \object{[OMN2000] LS1}  occuring at least $\sim$3Myr ago. 
Precise limits depend on both distance and stellar rotation, which is currently unquantifiable, but  both  theoretical and 
observational considerations  
suggest 3-6Myr  for a distance of  6kpc and $>$10Myr for 2kpc, if such a low luminosity solution is tenable (with the intermediate value of 3.4kpc implying a range of $\sim$5-10Myr).

In an analagous manner, the numbers and ratios of OB supergiants to  Wolf Rayets  and cool evolved stars 
such as Yellow Hypergiants and  RSGs also  potentially constrain recent ($\sim$20Myr) SF within W51.
Hadfield et al. (\cite{hadfield}) showed that hot post-MS 
evolutionary phases may be identified by an IR excess due to stellar winds.
While a number of stars with such excesses are found within the W51 complex (van Dyk et al., priv. comm. 2008), 
photometric data alone do not allow for the discrimination of their precise  evolutionary state, which is 
necessary  to constrain the SF history\footnote{There exists the potential for the miss-identification of 
pre-MS objects as post-MS stars, with  Hadfield   et al. (\cite{hadfield}) finding a large number of B[e] stars 
using their IR selection criteria; a heterogeneous  classification containing stars at very different evolutionary 
stages (Lamers et al. \cite{lamers}).}. However,  cool evolved stars such as RSGs are readily identifiable 
even  at a distance of 6kpc and A$_v=25$mag (Davies et al. \cite{davies08}). Examination of the 2MASS
data for the complex reveals only a single  bright candidate (2MASS J19225290+1411210; J=7.21, H=5.50, K=4.63), with colours 
consistent with a moderately reddened M star (A$_v \sim8$; colours from Elias et al. \cite{elias}), for which we 
infer log(L/L$_{\odot}$)$\sim$5.5 (5.0) for a putative spectral type of M0 I 
(M5 I; bolometric corrections  from Levesque et al. \cite{levesque}) at a distance of 6kpc
and hence an age in the range of $\sim$10-14Myr  (e.g. Davies et al. \cite{davies08}), noting that the comparatively low 
extinction could be the result of clearing of the local ISM by precursor O stars (as suggested by the anonymous referee).

\begin{figure}
\resizebox{\hsize}{!}{\includegraphics[angle=0]{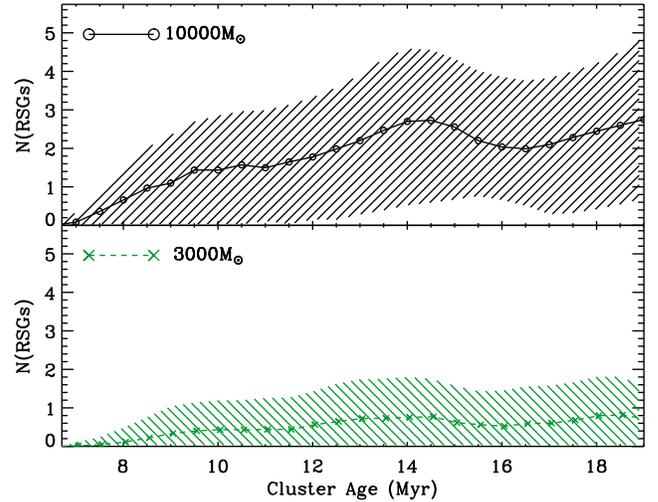}}
\caption{Plot showing the mean number of RSGs expected for clusters with masses of 3 \& 
10$\times$10$^3$M$_{\odot}$ as a  function of cluster age, the hashed regions indicating the formal 1$\sigma$ range
derived from the Monte Carlo simulations. By comparison, current SF within W51 
is forming clusters in the 
2-10$\times$10$^3$M$_{\odot}$ range, and the Orion Nebula cluster is $\sim$2000M$_{\odot}$ (Sect. 4).
The dip between 14-20Myr is caused by the onset of a blue loop for a restricted mass range of progenitors.}
\end{figure}

\begin{figure}
\resizebox{\hsize}{!}{\includegraphics[angle=0]{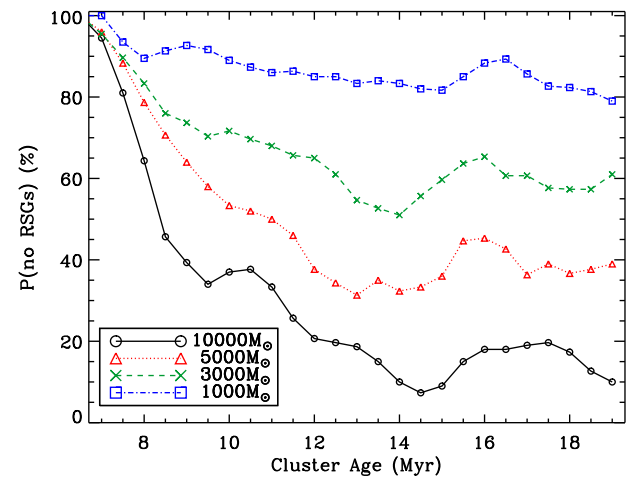}}
\caption{Plot showing the probability that a cluster of given age and mass has {\em no} RSGs present (Sect. 4).}
\end{figure}

Irrespective of the actual classification of 2MASS J19225290+1411210, a large population of  RSGs 
appears to be absent from W51. Following the methodology of  Davies et al. (\cite{davies08} and refs. therein) we may
 utilise this observation to interpret the prior SF history. We built synthetic clusters of differing masses in the  
1-40,000M$_{\odot}$ range, populating them according to a Kroupa type Initial Mass Function (Kroupa 
\cite{kroupa})\footnote{Previous simulations of RSGC1-3 (Davies et al. \cite{davies07}, \cite{davies08}, Clark et al. 
\cite{clark09}) had assumed a Salpeter IMF - for a given population of RSGs the Kroupa IMF yeilds more low mass stars, 
resulting in an increase in cluster mass by $\sim$30\% over the estimates given in these works.}, and 
evolving stars according to the  rotating (v$_{rot}$=300kms$^{-1}$), solar metalicity 
  evolutionary tracks of  Meynet \& Maeder (\cite{meynet}).
RSGs were classified as those stars with log(L/L$_{\odot}$) 
$>$ 4.5 and T $<$ 4500K, and 200 trials  per age interval were employed 
 to reduce random noise. Results from these simulations, in terms of both 
the mean number of RSGs expected and the probability that none will be present, as a function of cluster mass 
and age, are presented in Figs. 3 \& 4.

These results clearly demonstrate that, given the brevity of the RSG phase, statistically significant 
conclusions may only been drawn for very massive ($>>10^4$M$_{\odot}$) clusters which, by extention, may 
host correspondingly rich populations of RSGs. Indeed, RSGs are so rare at  ages of $<$8Myr that no meaningful  
constraints may be placed  on the properties of individual host clusters, even for  such 
extreme  masses.  These uncertainties are further exacerbated  in the event that  constraints on the age 
of the  RSG population  are unavailable,  which would occur if no estimate of the stellar luminosity were available. 
Since this would be the case for  2MASS  J19225290+1411210 {\em  if it were a RSG},  we may  only infer that any putative
natal cluster must be greater than 5~Myr in age. 
However, we may compare the results of the Monte Carlo simulations to the properties of the ensemble of clusters 
identified by  Nanda Kumar et al. ({\cite{nanda})   to investigate  whether the global rate of SF in the W51 complex was 
larger or smaller in the past.   Nanda Kumar et al. ({\cite{nanda}; Fig. 1) reported the discovery of 5 massive 
(2-10$\times 10^3$M$_{\odot}$), young (0.7-3Myr) clusters in the 4 fields they surveyed. Given the results presented in Fig. 4,
 if these clusters were to be observed at an  age of  9(20)~Myr we would expect a total of  3(20) RSGs to be present, 
with only a  $<$5\% chance that none would be present at either epoch. We therefore conclude that the current 
SF  differs from past ($>$9Myr) epochs in the sense that greater numbers of massive clusters are currently being formed.

While we defer a detailed description of the Spitzer data to a future paper, we note that the
 morphology of the emission also casts doubt on the hypothesis of OMN2000 that SF in their 
Region 1 subsequently triggered similar activity in their
Region 3 (which contains the massive proto-cluster IRS2; Figueredo et al. \cite{figueredo}, Barbosa et al. \cite{barbosa}). Under 
such a scenario we  might expect to see a mid-IR morphology similar to that of the G305 SF
complex  (Clark \& Porter \cite{clark04}), with IRS2 
residing on the periphery of a large wind blown cavity, which would be readily visible in e.g. 
Spitzer data (Churchwell et al. \cite{churchwell}). 
While such structures are found within Region 1 of OMN2000 (Fig. 1), IRS2 clearly resides beyond 
their boundaries, implying that it formed independently of this activity.

Therefore,  in summary we are able to conclude that SF activity  within the W51 complex resulting in 
the production of massive O stars is ongoing and has proceeded at 
multiple sites throughout the cloud over  {\em at least} the last $\sim$3Myr, with no current evidence 
for widespread {\em internal} sequential triggering, although the simultaneity of these events 
does suggest a large scale external trigger (c.f Nanda Kumar et al. \cite{nanda}). 

Recent observations of external galaxies such as M51 have revealed that  SF appears to yield complexes of star clusters with a 
range of ages  ($\sim$10Myr; Bastian et al. \cite{bastian}) and it appears likely that W51, along 
with other complexes such as G305 (Clark \&  Porter \cite{clark04}) and the Carina Nebula (Smith \& Brooks 
\cite{smith}) are Galactic analogues.
However, despite the physical similarities demonstrated by these complexes - a heirarchical  
distribution of star formation on multiple spatial scales  which likely  reflect the fractal nature 
of their natal GMCs (Elmegreen \cite{elmegreen}) -   differences between the locations of distinct 
stellar populations are apparent. For example  the  extragalactic Giant HII region NGC2403-I consists 
of  a halo of 7-10Myr old  RSGs surrounding  a population of significantly younger   (2-6Myr) massive  
stellar clusters   (Drissen et al. \cite{drissen99}). Conversely 30 Dor and the G305 complex  
 comprise dense central clusters (2-5Myr) which are triggering  new waves of SF ($\leq$1Myr) 
on the periphery of wind blown bubbles (Walborn et al. \cite{walborn}, Clark \& Porter \cite{clark04}), while 
we currently find  no evidence for spatially segregated, sequentially triggered  star formation in W51.

\section{Conclusions}

Utilising new near-IR spectroscopic observations  we find that \object{[OMN2000] LS1} is a massive evolved 
star best classified as an extreme P Cygni B supergiant. Our non-LTE analysis of the star suggests a 
significant downwards revision in stellar temperature, luminosity and initial mass when compared to the values presented by 
OMN2000. Assuming the spectroscopically determined distance of 2kpc to W51 (Figueredo et al. \cite{figueredo})
 results in  a luminosity and  progenitor mass significantly lower than expected for such stars on both 
observational  and theoretical grounds. In contrast, a distance of 6kpc - representative of both kinematic
and parallactic estimates - results in a luminosity of 
log(L/L$_{\odot}$) $\sim$5.75, an intial  mass of order $\sim$40M$_{\odot}$, a mass loss rate of 
6.6$\times10^{-5}$M$_{\odot}$yr$^{-1}$ and and elevated He/H ratio; entirely consistent with quantitative analyses
of other P Cygni supergiants, LBVs and the closely related WN9-11 stars. Given a current  empirical  minimum luminosity of
 log(L/L$_{\odot}$) $\sim$5.3 for the P Cygni supergiants, we suggest a corresponding minimum  distance of $\sim$3.4kpc
 for \object{[OMN2000] LS1}.

Nevertheless, for either  distance, 
this result demonstrates that massive SF  in W51 has been underway for a {\em minimum} of 3Myr 
and is still ongoing (e.g. Figueredo et al. \cite{figueredo}). 
However,  the lack of a significant population of  RSGs -   if indeed {\em any} are present - 
within the complex  suggests that the formation of star clusters differed in the 
past, such that the massive clusters being formed now (Nanda Kumar et al. \cite{nanda})  were not forming
 $\geq$9Myr ago.

 The morphology of the mid-IR emission  surrouding  \object{[OMN2000] LS1} casts doubt upon the
 hypothesis of OMN2000 that sequential, internally triggered SF has occurred within W51, but is 
consistent with the suggestion of Nanda Kumar et al. (\cite{nanda}) of activity at multiple sites initiated by an external 
trigger.  In this respect W51 differs from other  complexes such as G305 and the Carina nebula, where ongoing SF 
triggered by the first generation of stars appears to be occuring on the periphery of a wind blown cavity in the GMC.

 Nevertheless, all three regions appear to be  Galactic counterparts to the star forming complexes identified 
in external galaxeis such as M51 (Bastian et al. \cite{bastian}) and  NGC2403 (Drissen et al. \cite{drissen99}), 
which are characterised by SF on multiple spatial scales and over a comparatively  short ($\sim$10Myr) period of time. 
While the global, heirarchical  properties of such complexes likely represent the initial 
conditions of the natal GMCs (Elmegreen \cite{elmegreen}), the relative magnitudes of, and 
interplay between,   internal 
(feedback from OB stars \& SNe) and external triggering processes 
(SNe, passage of galactic density waves and  interaction with external 
galaxies) likely play a key role in determining the detailed SF history 
and morphology of individual examples.

\begin{acknowledgements}
JSC is funded by an RCUK fellowship. 
F. Najarro acknowledges Spanish grants AYA2008-06166-C03-02 and GTC
  Consolider.
IRMOS was developed with the generous support of the
Space Telescope Science Institute, the James Webb
Space Telescope, the NASA Goddard Space Flight Center, 
the Kitt Peak National Observatory. The National Optical Astronomy Observatory (NOAO) consists 
of Kitt Peak National Observatory near Tucson, Arizona, Cerro Tololo Inter-American Observatory 
near La Serena, Chile, and the NOAO Gemini Science Center. NOAO is operated by the Association 
of Universities for Research in Astronomy (AURA) under a cooperative agreement with the National 
Science Foundation.

\end{acknowledgements}

\end{document}